\documentstyle[12pt]{article}
\begin{document}
 \title{4-Spinors and 5D Spacetime}
 \author{Francesco Antonuccio\footnote{The author is trained as a theoretical physicist and is currently a portfolio manager in a London hedge fund.} \\
 London, United Kingdom\\
\texttt{f\_antonuccio@yahoo.co.uk}}
 \renewcommand{\today}{July 10th, 2013}
 \maketitle
 \abstract 
We revisit the subject exploring maps from the space of 4-spinors to 3+1 space-time that commute with the Lorentz transformation. All known mappings have a natural embedding in a higher five dimensional spacetime, and can be succinctly expressed as products of quaternions, or split-quaternions, depending on the signature of the embedding 5D spacetime. It is in this sense that we may view the geometry of 4-spinors as being related to the `square root' of five dimensional spacetime. In particular, a point in 5D spacetime may be identified with a corresponding 4-spinor that is uniquely determined up to a quaternionic - or split-quaternionic - phase.   \\
\section {Introduction}
We revisit the idea that one can directly map the space of 4-spinors onto 3+1 spacetime in a way that preserves the Lorentz transformation across the two different representations. Formally, these maps commute with the Lorentz transformation. This was first discussed in \cite{AntonuccioSpinorPaperII} where we provided details of two such mappings.
There are now known to be four distinct mappings, each of which has a natural embedding in a higher five dimensional spacetime. In particular, one map has a natural embedding in a spacetime with signature $(+,-,-,-,-)$, while the three remaining maps are naturally embedded in a spacetime with signature $(-,+,+,+,-)$.  Writing down the explicit form of these four maps will be the main goal of this paper.

All of these maps admit a simple realisation in terms of products of quaternions, or split quaternions, depending on the signature of the embedding five dimensional spacetime. These results allow us to find the missing degrees of freedom when attempting to loosely `identify' a 4-spinor with a point in the embedding 5D spacetime; namely, the spinor is defined uniquely up to a (three dimensional) quaternionic -- or split-quaternionic -- phase. A formal study of quaternions and split-quaternions and their connection to physics in four dimensional spacetime can be found in a series of papers by Frenkel and Libine (see \cite{Frenkel}, \cite{FrenkelII}). 

The paper is organised as follows: In Section \ref{sec:HLT}, we summarise the transformation properties of hyperbolic 4-spinors under Lorentz transformations in 3+1 dimensions. It has been shown that the transformation properties of the eight real components of complex 4-spinors and hyperbolic 4-spinors are identical after a suitable identification of real components is made between representations \cite{AntonuccioSpinorPaperI}. The hyperbolic 4-spinor is just the Dirac 4-spinor in disguise. Consequently,  the choice of representation (complex or hyperbolic) is, in principle, arbitrary, and does not change the final results of this paper. Nevertheless, at the risk of introducing largely unfamiliar concepts to the reader, we choose to focus on the hyperbolic representation of the Lorentz group. One obvious advantage is that Lorentz invariance in spinor space becomes  manifest when working with hyperbolic numbers ${\bf D}$. For example, the Lorentz group which acts on a hyperbolic 4-spinor $\Psi \in {\bf D}^4$ turns out to be a subgroup of the fifteen dimensional special unitary group SU$(4,{\bf D})$, and so the real-valued quantity $|\Psi|^2 = \Psi^{\dag} \Psi$ is automatically a Lorentz invariant scalar. For a simple and concise introduction to the algebra of hyperbolic numbers and the corresponding unitary groups, see \cite{AntonuccioSpinorPaperII}.

In Section \ref{sec:MapSpinorsToSpacetime} we explicitly write down four distinct maps from spinor space to 3+1 spacetime that commute with the Lorentz transformation. Equivalently, these maps preserve the Lorentz transformation properties of components across different representations. These maps take an especially concise form when represented in terms of quaternions, and make manifest the presence of an additional fifth spacetime coordinate. Physically, the fifth coordinate is hidden from view since it transforms as a scalar under these same 3+1 Lorentz transformations and so does not mix with the usual spacetime coordinates. 

We summarise our results and discuss applications in Section \ref{sec:Conclusions}.

\section{Hyperbolic 4-Spinors }
\label{sec:HLT}

Since hyperbolic representations of the Lorentz group may be unfamiliar to the reader, we will provide an introduction to the subject, culminating in the transformation properties of hyperbolic 4-spinors. In what follows, a hyperbolic number $w \in {\bf D}$ can always be written in the form 
\begin{equation}
w = a +{\rm j} b
\end{equation}
where $a$ and $b$ are real and ${\rm j}$ is a commuting variable satisfying
\begin{equation}
{\rm j}^2 = 1.
\end{equation}
For more details, the reader is referred to the general literature \cite{scn}.

\subsection{The Standard Lorentz Group on 3+1 Spacetime}
Given coordinates $(T,X,Y,Z)$ in 3+1 spacetime with metric signature $(+---)$, a Lorentz transformation on these coordinates corresponds to any  linear map (i.e. $4 \times 4$ matrix $L$)  acting on the four coordinates of spacetime,
\begin{equation}
\label{eq: lorentzmap}
\begin{array}{ccc}
\left(
\begin{array}{c}
T \\
X \\
Y \\
Z
\end{array}
\right)

& \rightarrow &

L \cdot \left(
\begin{array}{c}
T \\
X \\
Y \\
Z
\end{array}
\right),

\end{array}
\end{equation}
which preserves the quantity
\begin{equation}
\label{eq: 4Dlorentzscalar}
T^2-X^2-Y^2-Z^2.
\end{equation} 
These conditions imply that if $L$ is continuoulsy connected to the identity, it must take the form
\begin{equation}
\label{eq: realLorentztransform}
L = e^{\alpha_1 E_1+\alpha_2 E_2+\alpha_3 E_3+\beta_1 F_1+\beta_2 F_2 +\beta_3 F_3 }
\end{equation} 
where the $\alpha_i$ and $\beta_i$ are real numbers and $E_i$,$F_i$ are real matrices defined as follows:
\begin{equation}
\begin{array}{ccc}
E_1 = \left(
\begin{array}{cccc}
0 & 0 & 1 & 0 \\
0 & 0 & 0 & 0 \\
1 & 0 & 0 & 0 \\
0 & 0 & 0 & 0
\end{array}
\right)
&
E_2 = \left(
\begin{array}{cccc}
0 & 0 & 0 & 0 \\
0 & 0 & 0 & 0 \\
0 & 0 & 0 & 1 \\
0 & 0 & -1 & 0
\end{array}
\right)
&
E_3 = \left(
\begin{array}{cccc}
0 & 0 & 0 & 1 \\
0 & 0 & 0 & 0 \\
0 & 0 & 0 & 0 \\
1 & 0 & 0 & 0
\end{array}
\right)
\\
\\
F_1 = \left(
\begin{array}{cccc}
0 & 0 & 0 & 0 \\
0 & 0 & 0 & -1 \\
0 & 0 & 0 & 0 \\
0 & 1 & 0 & 0
\end{array}
\right)
 & 
F_2 = \left(
\begin{array}{cccc}
0 & -1 & 0 & 0 \\
-1 & 0 & 0 & 0 \\
0 & 0 & 0 & 0 \\
0 & 0 & 0 & 0
\end{array}
\right)
& 
F_3 = \left(
\begin{array}{cccc}
0 & 0 & 0 & 0 \\
0 & 0 & 1 & 0 \\
0 & -1 & 0 & 0 \\
0 & 0 & 0 & 0
\end{array}
\right)

\end{array}
\end{equation}
From a physical point of view, $E_1$, $F_2$ and $E_3$ correspond to Lorentz boosts parallel to the three spatial directions, while $F_1$, $E_2$ and $F_3$ correspond to rotations about the three spatial coordinate axes. It is in this sense that there are six generators for the Lorentz group on 3+1 spacetime.  

By direct substitution, one can check that these six generators of the Lorentz group satisfy the following commutation relations:
\begin{equation}
\begin{array}{llll}
\label{eq: 4Dliealgebra}
[E_1,E_2] = E_3  &  [F_1,F_2] = -E_3  &  [E_1,F_2] = F_3  &  [F_1,E_2] = F_3  \\
\left[E_2,E_3 \right] = E_1  &  \left[F_2,F_3 \right] = -E_1  &  \left[ E_2,F_3 \right] = F_1  &  \left[ F_2,E_3 \right] = F_1 \\
\left[ E_3,E_1 \right] = -E_2  &  \left[ F_3,F_1 \right] = E_2  &  \left[ E_3,F_1 \right] = -F_2  &  \left[ F_3,E_1 \right] = -F_2 
\end{array}
\end{equation}

All other commutators vanish. Abstractly, these relations define the Lie Algebra of the Lorentz group on 3+1 spacetime.

If we now view the real coeeficients $\alpha_i, \beta_i$ appearing in (\ref{eq: realLorentztransform}) as infinitesimally small, then the transformation (\ref{eq: lorentzmap}) on the 3+1 spacetime coordinates $(T,X,Y,Z)$ takes the following explicit form:
\begin{equation}
\label{eq: x0map}
T \rightarrow T -\beta_2 X+\alpha_1 Y +\alpha_3 Z
\end{equation}
\begin{equation}
X \rightarrow X -\beta_2 T+\beta_3 Y -\beta_1 Z
\end{equation}
\begin{equation}
Y \rightarrow Y +\alpha_1 T-\beta_3 X -\alpha_2 Z
\end{equation}
\begin{equation}
\label{eq: x3map}
Z \rightarrow Z +\alpha_3 T+\beta_1 X -\alpha_2 Y
\end{equation}
where we have dropped all terms beyond the linear approximation.

In the next Section, we find another representation of the Lorentz Lie algebra (\ref{eq: 4Dliealgebra}) in terms of $4 \times 4$ anti-Hermitian matrices defined over the hyperbolic numbers. 

\subsection{Hyperbolic Representation of the 3+1 Lorentz Group}
\label{lorentzinvariance}
Our objective in this section is to find $4 \times 4$ anti-Hermitian matrices over ${\bf D}$ that satisfy the Lorentz Lie algebra (\ref{eq: 4Dliealgebra}). We begin by defining three $2 \times 2$ anti-Hermitian matrices $\tau_1$, $\tau_2$ and $\tau_3$ as follows:
\begin{equation}
\begin{array}{ccc}
\tau_1 = \frac{1}{2}
\left(
\begin{array}{cc}
0 & {\rm j} \\
{\rm j} & 0
\end{array}
\right)
&
\tau_2 = \frac{1}{2}
\left(
\begin{array}{cc}
0 & -1 \\
1 & 0
\end{array}
\right)
&
\tau_3 = \frac{1}{2}
\left(
\begin{array}{cc}
{\rm j} & 0 \\
0 & -{\rm j}
\end{array}
\right)
\end{array}
\end{equation}
It follows that these matrices satisfy the following commutation relations:
\begin{equation}
\begin{array}{ccc}
\left[ \tau_1, \tau_2 \right] = \tau_3  &  \left[ \tau_2, \tau_3 \right] = \tau_1 & \left[ \tau_3, \tau_1 \right] = -\tau_2 .
\end{array}
\end{equation}
We are now ready to define the $4 \times 4$ matrices $E_i$ and $F_i$ that will form the basis of our hyperbolic representation:
\begin{equation}
\label{eq: hyperbolicrep}
\begin{array}{ccc}
E_1 = \left(
\begin{array}{cc}
\tau_1 & 0 \\
0 & \tau_1
\end{array}
\right)
&
E_2 = \left(
\begin{array}{cc}
\tau_2 & 0 \\
0 & \tau_2
\end{array}
\right)
&
E_3 = \left(
\begin{array}{cc}
\tau_3 & 0 \\
0 & \tau_3
\end{array}
\right)
\\ \\
F_1 = {\rm j} \left(
\begin{array}{cc}
 0 & \tau_1 \\
-\tau_1 & 0
\end{array}
\right)
&
F_2 = {\rm j} \left(
\begin{array}{cc}
 0 & \tau_2 \\
-\tau_2 & 0
\end{array}
\right)
&
F_3 = {\rm j} \left(
\begin{array}{cc}
 0 & \tau_3 \\
-\tau_3 & 0
\end{array}
\right)
\end{array}
\end{equation}
It is straightforward to check that the matrices  $E_i$ and $F_i$ defined above are indeed anti-Hermitian with respect to ${\bf D}$  (i.e. $E_{i}^{\dag} = -E_i$ and $F_{i}^{\dag} = -F_i$), and satisfy the 3+1 Lorentz Lie algebra of commutation relations (\ref{eq: 4Dliealgebra}). 
Consequently, a Lorentz transformation $L$ in the representation specified by the generating matrices (\ref{eq: hyperbolicrep}) takes the form 
\begin{equation}
\label{eq: realLorentztransformII}
L = e^{\alpha_1 E_1+\alpha_2 E_2+\alpha_3 E_3+\beta_1 F_1+\beta_2 F_2 +\beta_3 F_3 }
\end{equation} 
where the $\alpha_i$ and $\beta_i$ are real numbers as before, but $L$ now acts on 4-component hyperbolic spinors $\Psi \in {\bf D}^4$:
\begin{equation}
\label{eq: explicithyperbolictransform}
 \left(
 \begin{array}{c}
a_1+ \rm{j} b_1 \\
a_2+ \rm{j} b_2 \\
a_3+ \rm{j} b_3 \\
a_4+ \rm{j} b_4
\end{array}
\right) \rightarrow
L \cdot 
 \left(
 \begin{array}{c}
a_1+ \rm{j} b_1 \\
a_2+ \rm{j} b_2 \\
a_3+ \rm{j} b_3 \\
a_4+ \rm{j} b_4
\end{array}
\right) .
\end{equation}
The components $a_i$ and $b_i$ appearing above are real numbers. Since the $E_i$ and $F_i$ appearing in  (\ref{eq: realLorentztransformII}) are anti-Hermitian, then 
\begin{equation}
L^\dag = e^{-\left(\alpha_1 E_1+\alpha_2 E_2+\alpha_3 E_3+\beta_1 F_1+\beta_2 F_2 +\beta_3 F_3\right) } =L^{-1}.
\end{equation}
In other words $L^{\dag}$ is the inverse of $L$:
\begin{equation}
\label{eq: inversecondition}
L^{\dag} L = L L^{\dag} = 1.
\end{equation}
Moreover, the generating matrices $E_i$ and $F_i$  are traceless, so we have the additional property
\begin{equation}
\label{eq: detcondition}
{\rm det} L = 1.
\end{equation}
It follows from (\ref{eq: inversecondition}) and  (\ref{eq: detcondition}) above that the Lorentz transformation $L$ defined by  (\ref{eq: realLorentztransformII}) is an element of the special hyperbolic unitary group SU$(4,{\bf D})$. This group manifold actually has fifteen generators in its Lie algebra, and so we conclude that the Lorentz group on 3+1 spacetime is a subgroup of SU$(4,{\bf D})$. 

If we assume the real coefficients $\alpha_i$ and $\beta_i$ appearing in (\ref{eq: realLorentztransformII})  are infinitesimally small, then up to linear order in these coefficients, we may write the Lorentz transformation  (\ref{eq: explicithyperbolictransform}) explicitly in terms of the eight real components $a_i$ and $b_i$:
\begin{equation}
\label{eq: a1map}
a_1 \rightarrow a_1 +\frac{1}{2} \left(
\alpha_1 b_2 -\alpha_2 a_2 + \alpha_3 b_1 + \beta_1 a_4 - \beta_2 b_4 + \beta_3 a_3
\right)
\end{equation}
\begin{equation}
a_2 \rightarrow a_2 +\frac{1}{2} \left(
\alpha_1 b_1 +\alpha_2 a_1 - \alpha_3 b_2 + \beta_1 a_3 + \beta_2 b_3 - \beta_3 a_4
\right)
\end{equation}
\begin{equation}
a_3 \rightarrow a_3 +\frac{1}{2} \left(
\alpha_1 b_4 -\alpha_2 a_4 + \alpha_3 b_3 - \beta_1 a_2 + \beta_2 b_2 - \beta_3 a_1
\right)
\end{equation}
\begin{equation}
a_4 \rightarrow a_4 +\frac{1}{2} \left(
\alpha_1 b_3 +\alpha_2 a_3 -\alpha_3 b_4 - \beta_1 a_1 - \beta_2 b_1 + \beta_3 a_2
\right)
\end{equation}
\begin{equation}
b_1 \rightarrow b_1 +\frac{1}{2} \left(
\alpha_1 a_2 -\alpha_2 b_2 + \alpha_3 a_1 + \beta_1 b_4 - \beta_2 a_4 + \beta_3 b_3
\right)
\end{equation}
\begin{equation}
b_2 \rightarrow b_2 +\frac{1}{2} \left(
\alpha_1 a_1 +\alpha_2 b_1 - \alpha_3 a_2 + \beta_1 b_3 + \beta_2 a_3 - \beta_3 b_4
\right)
\end{equation}
\begin{equation}
b_3 \rightarrow b_3 +\frac{1}{2} \left(
\alpha_1 a_4 -\alpha_2 b_4 + \alpha_3 a_3 - \beta_1 b_2 + \beta_2 a_2 - \beta_3 b_1
\right)
\end{equation}
\begin{equation}
\label{eq: b4map}
b_4 \rightarrow b_4 +\frac{1}{2} \left(
\alpha_1 a_3 +\alpha_2 b_3 - \alpha_3 a_4 - \beta_1 b_1 - \beta_2 a_1 + \beta_3 b_2
\right)
\end{equation}

It was pointed out in \cite{AntonuccioSpinorPaperI} that these transformations are in one-to-one correspondence with the transformations acting on the eight real components of a Dirac (i.e. complex) 4-spinor, and so the hyperbolic 4-spinor is just the Dirac 4-spinor in disguise!

The infinitesimal Lorentz transformations on real spacetime points given by (\ref{eq: x0map})-(\ref{eq: x3map}), and on hyperbolic spinors specified by (\ref{eq: a1map})-(\ref{eq: b4map})  above will be relevant for the next section. In particular, we show there exist four distinct projections that map the spinor components $a_i$ and $b_i$ to a real spacetime point $(T,X,Y,Z)$  that respects simultaneously the relations  (\ref{eq: x0map})-(\ref{eq: x3map}) and (\ref{eq: a1map})-(\ref{eq: b4map}) under an infinitesimal Lorentz transformation. Formally, there exists a projection map that commutes with the Lorentz transformation.

\section{Mappings from Spinors to Spacetime}
\label{sec:MapSpinorsToSpacetime}
Suppose we are given a hyperbolic spinor $\Psi \in {\bf D}^4$. Then we may write 
\begin{equation}
\label{eq: spinor}
\Psi = \left(
 \begin{array}{c}
a_1+ \rm{j} b_1 \\
a_2+ \rm{j} b_2 \\
a_3+ \rm{j} b_3 \\
a_4+ \rm{j} b_4
\end{array}
\right),
\end{equation}
where the eight components $a_i$ and $b_i$ are real numbers. Our objective is to find mappings of the form 
\begin{equation}
\Psi \rightarrow (T,X,Y,Z)
\end{equation}
which give rise to the 3+1 spacetime Lorentz transformation (\ref{eq: x0map})-(\ref{eq: x3map}) whenever we transform the eight real spinor components of $\Psi$ in accordance with (\ref{eq: a1map})-(\ref{eq: b4map}). The four known mappings that satisfy this condition take a particularly simple form when expressed in terms of either quaternionic or split-quaternionic variables, and will be presented in Sections \ref{sec:MapSpinorsToSpacetimeI} - \ref{sec:MapSpinorsToSpacetimeIV}. However, for completeness, we shall first provide basic definitions for quaternions and split-quaternions \cite{sqn} next.

\subsection{Quaternions and Split-Quaternions}
A quaternion $q \in {\bf H}$ is any number of the form
\begin{equation}
\label{eq:qdefn}
q = z_1 + {\rm I} z_2
\end{equation}
where $z_1$ and $z_2$ are complex numbers, and ${\rm I}$ is a non-commuting variable satisfying
\begin{equation}
{\rm I}^2 = -1
\end{equation}
and 
\begin{equation}
{\rm I}z= {\overline z} {\rm I}.
\end{equation}
This number system is associative and distributive, but in general non-commutative. If we define the conjugate $q^{\dag}$ by writing
\begin{equation}
q^{\dag} \equiv {\overline z}_1- {\rm I} z_2,
\end{equation}
then
\begin{equation}
q^{\dag} q = q q^{\dag} = |z_1|^2 +|z_2|^2.
\end{equation}
It is customary to define the modulus squared by writing $|q|^2 = q^{\dag} q $ for any quaternion $q$. Note that if $|q|^2 \neq 0$ then 
\begin{equation}
q^{-1} \equiv \frac{q^{\dag}}{|q|^2}
\end{equation}
is a well defined and unique inverse for $q$.

Split-quaternions can be defined in an analogous way; the only difference is that we replace the two complex numbers $z_1$ and $z_2$ appearing in (\ref{eq:qdefn}) with two hyperbolic numbers $w_1$ and $w_2$. In particular, we say that $p \in {\bf P}$ is a split-quaternion if it has the form
\begin{equation}
\label{eq:pdefn}
p = w_1 + {\rm I} w_2
\end{equation}
where $w_1$ and $w_2$ are hyperbolic (i.e. split-complex) numbers, and ${\rm I}$ is a non-commuting variable satisfying
\begin{equation}
{\rm I}^2 = -1
\end{equation}
and 
\begin{equation}
{\rm I}w= {\overline w} {\rm I}.
\end{equation}
This number system is also associative and distributive, and in general non-commutative. If we define the conjugate $p^{\dag}$ by writing
\begin{equation}
p^{\dag} \equiv {\overline w}_1- {\rm I} w_2,
\end{equation}
then
\begin{equation}
p^{\dag} p = p p^{\dag} = |w_1|^2 +|w_2|^2.
\end{equation}
In what follows, we define a modulus squared for split-quaternions by writing $|p|^2 = p^{\dag} p $ for any split-quaternion $p$. Unlike the case for quaternions, the modulus squared of a split-quaternion can be negative.

Finally, if  $|p|^2 \neq 0$ then 
\begin{equation}
p^{-1} \equiv \frac{p^{\dag}}{|p|^2}
\end{equation}
is a well defined and unique inverse for $p$. 

Equipped with these definitions, we are now ready to define our four projection maps from spinor space to spacetime.

\subsection{Map I }
\label{sec:MapSpinorsToSpacetimeI}
For a given hyperbolic 4-spinor $\Psi$ with eight real components $a_i,b_i, ( i=1,\dots,4)$, as specified by definition (\ref{eq: spinor}), we define the following pair of quaternions:
\begin{equation}
\begin{array}{l}
q_1 = (a_1+{\rm i} a_4)+{\rm I}(a_2-{\rm i}a_3) \\
q_2 = (b_1+{\rm i} b_4)+{\rm I}(b_2+{\rm i}b_3) \\
\end{array}
\end{equation}
These two quaternions encode all the degrees of freedom of the spinor $\Psi$. Using these definitions, we can map the spinor  $\Psi$ to a point   $(T,X,Y,Z,U)$ in 5D spacetime by writing:
\begin{equation}
\label{eq:mapI}
\begin{array}{rcl}
T & = & \frac{1}{2} (|q_1|^2+|q_2|^2) \\
(Z+{\rm i}X)+ {\rm I} (Y + {\rm i} U) & = & q_1 \cdot q_2
\end{array}
\end{equation}
It is straightforward to check that under a 3+1 Lorentz transformation acting on the eight real spinor components specified by relations (\ref{eq: a1map})-(\ref{eq: b4map}), the mapping above guarantees that the spacetime variables $(T,X,Y,Z)$ transform in accordance with (\ref{eq: x0map})-(\ref{eq: x3map}), which is the usual Lorentz transformation acting on spacetime coordinates in 3+1 dimensions. The additional (fifth) spacetime coordinate $U$ appearing above is unavoidable in our quaternionic formalism, but is physically hidden from view since it transforms as a Lorentz scalar, and does not mix with the other spacetime coordinates.

We also deduce from (\ref{eq:mapI}) the following identity:
\begin{equation}
\label{eq:L1}
\left[\frac{1}{2}|\Psi|^2\right]^2  =\frac{1}{4}\left(|q_1|^2- |q_2|^2\right)^2 = T^2-X^2-Y^2-Z^2-U^2.
\end{equation}
It is in this sense that the mapping is naturally embedded in a 5D spacetime with signature $(+,-,-,-,-)$, where the additional spacetime coordinate $U$ is space-like.

Note that the map described here is not a bijection between spinors and 5D spacetime. The spinor $\Psi$ has eight real components, while the spacetime point has only five. Where are the three missing degrees of freedom? The quaternionic formalism we have introduced yields a simple answer to this question. Namley, if $u \in {\bf H}$ is a unit quaternion (i.e. $u^{\dag}u = 1$), then the transformation 
\begin{equation}
 \begin{array}{l}
q_1 \rightarrow q_1 \cdot u^{\dag} \\
q_2 \rightarrow u \cdot q_2 \\
\end{array}
\end{equation}
leaves the spacetime coordinates $(T,X,Y,Z,U)$ unchanged under the mapping (\ref{eq:mapI}). The unit quaternion is characterised by a three dimensional space of parameters, and is a natural generalisation of the one-dimensional complex phase. Hence points in 5D spacetime may be identified with a corresponding spinor modulo a quaternionic (i.e. three dimensional)  phase.

\subsection{Map II}
\label{sec:MapSpinorsToSpacetimeII}
Once again, assume we are given a hyperbolic 4-spinor $\Psi$ as defined by  (\ref{eq: spinor}). With the eight real components $a_i,b_i$ appearing in this definition, construct the following pair of split-quaternions:
\begin{equation}
\begin{array}{l}
p_1 = (a_1-{\rm j} b_1)+{\rm I}(-a_2+{\rm j}b_2) \\
p_2 = (a_3+{\rm j} b_3)+{\rm I}(-a_4+{\rm j}b_4) \\
\end{array}
\end{equation}
We now define a map that assigns to a given spinor $\Psi$ a point $(T,X,Y,Z,U)$ in 5D spacetime by writing:
\begin{equation}
\label{eq:mapII}
\begin{array}{rcl}
X & = & \frac{1}{2} (|p_1|^2-|p_2|^2) \\
(Y+{\rm j}U)+ {\rm I} (Z + {\rm j} T) & = & p_1 \cdot p_2
\end{array}
\end{equation}
Note that since $p_1,p_2$ are split-quaternions, the quantities $|p_1|^2$ and $|p_2|^2$ appearing above can be negative. As before, a straightforward calculation reveals that the above mapping commutes with the 3+1 spacetime Lorentz transformation, while the additional fifth spacetime coordinate $U$ acts as a Lorentz scalar. Moreover, one can show that equations (\ref{eq:mapII}) give rise to the following identity:
\begin{equation}
\label{eq:L2}
\left[\frac{1}{2}|\Psi|^2\right]^2  =\frac{1}{4}\left(|p_1|^2+|p_2|^2\right)^2 = - T^2+X^2+Y^2+Z^2-U^2.
\end{equation} 
We conclude that the embedding 5D spacetime for this map has signature $(-,+,+,+,-)$, and that the fifth spacetime coordinate $U$ is time-like.

Now if $v \in {\bf P}$ is a unit split-quaternion (i.e. $v^{\dag}v = 1$), then the transformation 
\begin{equation}
\label{eq:sqt}
 \begin{array}{l}
p_1 \rightarrow p_1 \cdot v^{\dag} \\
p_2 \rightarrow v \cdot p_2 \\
\end{array}
\end{equation}
leaves the spacetime coordinates $(T,X,Y,Z,U)$ unchanged under the mapping (\ref{eq:mapII}). Hence, under this mapping, one may identify a 5D spacetime point with a corresponding spinor, modulo a (three dimensional) split-quaternionic phase.

\subsection{Map III}
\label{sec:MapSpinorsToSpacetimeIII}
As above, given the eight real components $a_i,b_i$ appearing in the definition for the hyperbolic spinor $\Psi$, define a pair of split-quaternionic variables as follows:
\begin{equation}
\begin{array}{l}
p_1 = (a_1-{\rm j} b_4)+{\rm I}(a_4+{\rm j}b_1) \\
p_2 = (-a_2+{\rm j} b_3)+{\rm I}(-a_3+{\rm j}b_2) \\
\end{array}
\end{equation}
We now define a map that assigns to a given spinor $\Psi$ a point $(T,X,Y,Z,U)$ in 5D spacetime by writing:
\begin{equation}
\label{eq:mapIII}
\begin{array}{rcl}
Y & = & \frac{1}{2} (|p_1|^2-|p_2|^2) \\
(Z+{\rm j}U)+ {\rm I} (X + {\rm j} T) & = & p_1 \cdot p_2
\end{array}
\end{equation}
The above map can also be shown to commute with the 3+1 spacetime Lorentz transformation, and the additional fifth spacetime coordinate $U$ is a Lorentz invariant scalar. The relations (\ref{eq:mapIII}) also give rise to the following identity:
\begin{equation}
\label{eq:L3}
\left[\frac{1}{2}|\Psi|^2\right]^2  =\frac{1}{4}\left(|p_1|^2+|p_2|^2\right)^2 = - T^2+X^2+Y^2+Z^2-U^2.
\end{equation} 
We conclude that the embedding 5D spacetime for this map has signature $(-,+,+,+,-)$ since the additional spacetime coordinate $U$ is time-like. 

Finally, transformations of the form (\ref{eq:sqt}) leave the spacetime point $(T,X,Y,Z,U)$ invariant under the mapping (\ref{eq:mapIII}), and so we may identify points in this 5D spacetime with corresponding spinors modulo an ambiguity characterised by a three dimensional split-quaternionic phase.

\subsection{Map IV}
\label{sec:MapSpinorsToSpacetimeIV}
For our fourth map, the eight real components $a_i,b_i$ appearing in the definition for the hyperbolic spinor $\Psi$ are now used to construct the following pair of split-quaternions:
\begin{equation}
\begin{array}{l}
p_1 = (a_1+{\rm j} b_2)+{\rm I}(-a_3+{\rm j}b_4) \\
p_2 = (a_2+{\rm j} b_1)+{\rm I}(a_4+{\rm j}b_3) \\
\end{array}
\end{equation}

A map that assigns to a given spinor $\Psi$ a point $(T,X,Y,Z,U)$ in 5D spacetime is stated below:
\begin{equation}
\label{eq:mapIV}
\begin{array}{rcl}
Z & = & \frac{1}{2} (|p_1|^2-|p_2|^2) \\
(Y+{\rm j}T)+ {\rm I} (X + {\rm j} U) & = & p_1 \cdot p_2
\end{array}
\end{equation}
This map shares the same property as the three previous maps: 3+1 Lorentz transformations are preserved across the different representations. It is worth repeating how one may prove this: First, transform the eight real spinor components $a_i,b_i$ as specified by relations (\ref{eq: a1map})-(\ref{eq: b4map}), and then confirm that the spacetime variables $(T,X,Y,Z)$ as defined by (\ref{eq:mapIV}) transform in accordance with relations  (\ref{eq: x0map})-(\ref{eq: x3map}). One may also separately confirm that the additional spacetime coordinate $U$ is invariant under 3+1 Lorentz transformations.

The relations (\ref{eq:mapIV}) imply the following identity:
 \begin{equation}
\label{eq:L4}
\left[\frac{1}{2}|\Psi|^2\right]^2  =\frac{1}{4}\left(|p_1|^2+|p_2|^2\right)^2 = - T^2+X^2+Y^2+Z^2-U^2.
\end{equation} 
Hence, the embedding 5D spacetime for this map has signature $(-,+,+,+,-)$ and the additional spacetime coordinate $U$ is time-like. 

Invariance of the map  (\ref{eq:mapIV}) under transformations of the form (\ref{eq:sqt}) allow us to identify points in this 5D spacetime with corresponding spinor variables up to an ambiguity measured by a three dimensional split-quaternionic phase. 

\section{Concluding Remarks}
\label{sec:Conclusions}

We have demonstrated in this article that there are maps from the space of 4-spinors to 3+1 spacetime which commute with the Lorentz transformation. These maps have a concise  expression in terms of quaternionic - or split-quaternionic - variables, and are naturally embedded in a higher five dimensional spacetime. The additional fifth spacetime coordinate that emerges from our quaternionic formalism transforms as a Lorentz scalar, and so does not mix with the other spacetime coordinates.

Such maps allow us to identify points in a 5D spacetime with a corresponding spinor, with an ambiguity measured by a three dimensional quaternionic, or split quaternionic, phase.

As a final remark, we discuss future directions of our study. Note first that the eight real components of a given 4-spinor $\Psi$ may be encoded by a pair of split-quaternions $p_1$ and $p_2$. Also, equations (\ref{eq:L2}),(\ref{eq:L3}) and (\ref{eq:L4}) all satisfy the same form of identity:

\begin{equation}
\label{eq:L5}
\left[\frac{1}{2}|\Psi|^2\right]^2  = \frac{1}{4}\left(|p_1|^2 + |p_2|^2\right)^2 = - T^2+X^2+Y^2+Z^2-U^2.
\end{equation} 
 
Consequently, the group of linear transformations acting on the pair of split-quaternionic variables $p_1$, $p_2$ that preserves the scalar $|p_1|^2 + |p_2|^2$ certainly deserves more attention from a physics perspective. In particular, the group U$(2,{\bf P})$ of $2 \times 2$ unitary matrices over the split-quaternionic numbers appears to play a central role in the study of 4-spinors and associated five dimensional embeddings.  Considerable insight on related matters has already been achieved in a series of papers authored by Frenkel and Libine (see \cite{Frenkel}, \cite{FrenkelII}). We leave the investigation of the unitary split-quaternionic  group U$(2,{\bf P})$ and related manifolds over the split-quaternionic numbers to future work.

\end{document}